\documentclass[twocolumn,showpacs,preprintnumbers,amsmath,amssymb,superscriptaddress,floatfix,altaffilletter]{revtex4-1}

\usepackage{graphicx}
\usepackage{dcolumn}
\usepackage{bm}
\usepackage[figuresright]{rotating}
\usepackage{fancyhdr}
\usepackage{enumerate}
\usepackage{indentfirst}
\usepackage{xspace}
\usepackage{xfrac}
\usepackage{color}

\begin{document}

\title{\bf Measurement of inclusive $\pi^\circ$ production in the Charged-Current Interactions of Neutrinos in a 1.3-GeV wide band beam}


\newcommand{\BCN}{\affiliation{Institut de Fisica d'Altes Energies, Universitat Autonoma de Barcelona, E-08193 Bellaterra (Barcelona), Spain}}
\newcommand{\BU}{\affiliation{Department of Physics, Boston University, Boston, Massachusetts 02215, USA}}
\newcommand{\UBC}{\affiliation{Department of Physics \& Astronomy, University of British Columbia, Vancouver, British Columbia V6T 1Z1, Canada}}
\newcommand{\UCI}{\affiliation{Department of Physics and Astronomy, University of California, Irvine, Irvine, California 92697-4575, USA}}
\newcommand{\SACLAY}{\affiliation{Commissariat a l'Energie Atomique et aux Energies Alternatives, Centre de Saclay, IRFU/SPP, 91191 Gif-sur-Yvette, France}}
\newcommand{\CNU}{\affiliation{Department of Physics, Chonnam National University, Kwangju 500-757, Korea}}
\newcommand{\DU}{\affiliation{Department of Physics, Dongshin University, Naju 520-714, Korea}}
\newcommand{\DUKE}{\affiliation{Department of Physics, Duke University, Durham, North Carolina 27708, USA}}
\newcommand{\GENEVA}{\affiliation{DPNC, Section de Physique, University of Geneva, CH1211, Geneva 4, Switzerland}}
\newcommand{\UH}{\affiliation{Department of Physics and Astronomy, University of Hawaii, Honolulu, Hawaii 96822, USA}}
\newcommand{\KEK}{\affiliation{High Energy Accelerator Research Organization(KEK), Tsukuba, Ibaraki 305-0801, Japan}}
\newcommand{\HIR}{\affiliation{Graduate School of Advanced Sciences of Matter, Hiroshima University, Higashi-Hiroshima, Hiroshima 739-8530, Japan}}
\newcommand{\INR}{\affiliation{Institute for Nuclear Research, Moscow 117312, Russia}}
\newcommand{\KOBE}{\affiliation{Kobe University, Kobe, Hyogo 657-8501, Japan}}
\newcommand{\KOR}{\affiliation{Department of Physics, Korea University, Seoul 136-701, Korea}}
\newcommand{\KYO}{\affiliation{Department of Physics, Kyoto University, Kyoto 606-8502, Japan}}
\newcommand{\LSU}{\affiliation{Department of Physics and Astronomy, Louisiana State University, Baton Rouge, Louisiana 70803-4001, USA}}
\newcommand{\MIT}{\affiliation{Department of Physics, Massachusetts Institute of Technology, Cambridge, Massachusetts 02139, USA}}
\newcommand{\MIYAGI}{\affiliation{Department of Physics, Miyagi University of Education, Sendai 980-0845, Japan}}
\newcommand{\NIIGATA}{\affiliation{Department of Physics, Niigata University, Niigata, Niigata 950-2181, Japan}}
\newcommand{\OKAYAMA}{\affiliation{Department of Physics, Okayama University, Okayama, Okayama 700-8530, Japan}}
\newcommand{\OSAKA}{\affiliation{Department of Physics, Osaka University, Toyonaka, Osaka 560-0043, Japan}}
\newcommand{\ROME}{\affiliation{University of Rome Sapienza and INFN, I-00185 Rome, Italy}}
\newcommand{\SNU}{\affiliation{Department of Physics, Seoul National University, Seoul 151-747, Korea}}
\newcommand{\SOLTAN}{\affiliation{A.~Soltan Institute for Nuclear Studies, 00-681 Warsaw, Poland}}
\newcommand{\TOHOKU}{\affiliation{Research Center for Neutrino Science, Tohoku University, Sendai, Miyagi 980-8578, Japan}}
\newcommand{\SB}{\affiliation{Department of Physics and Astronomy, State University of New York, Stony Brook, New York 11794-3800, USA}}
\newcommand{\TUS}{\affiliation{Department of Physics, Tokyo University of Science, Noda, Chiba 278-0022, Japan}}
\newcommand{\KAM}{\affiliation{Kamioka Observatory, Institute for Cosmic Ray Research, University of Tokyo, Kamioka, Gifu 506-1205, Japan}}
\newcommand{\RCCN}{\affiliation{Research Center for Cosmic Neutrinos, Institute for Cosmic Ray Research, University of Tokyo, Kashiwa, Chiba 277-8582, Japan}}
\newcommand{\TRIUMF}{\affiliation{TRIUMF, Vancouver, British Columbia V6T 2A3, Canada}}
\newcommand{\VAL}{\affiliation{Instituto de F\'{i}sica Corpuscular, E-46071 Valencia, Spain}}
\newcommand{\UW}{\affiliation{Department of Physics, University of Washington, Seattle, Washington 98195-1560, USA}}
\newcommand{\WARSAW}{\affiliation{Institute of Experimental Physics, Warsaw University, 00-681 Warsaw, Poland}}

\BCN
\BU
\UBC
\UCI
\SACLAY
\CNU
\DU
\DUKE
\GENEVA
\UH
\KEK
\HIR
\INR
\KOBE
\KOR
\KYO
\LSU
\MIT
\MIYAGI
\NIIGATA
\OKAYAMA
\OSAKA
\ROME
\SNU
\SOLTAN
\SB
\TUS
\KAM
\RCCN
\TRIUMF
\VAL
\UW
\WARSAW

\author{C.~Mariani}\altaffiliation[Present Address: ]{Department of Physics, Columbia University, New York, NY 10027, USA}\ROME
\author{A.~Tornero-Lopez}\VAL
\author{J.~L.~Alcaraz}\BCN
\author{S.~Andringa}\BCN
\author{S.~Aoki}\KOBE
\author{Y.~Aoyama}\KOBE
\author{J.~Argyriades}\SACLAY
\author{K.~Asakura}\KOBE
\author{R.~Ashie}\KAM
\author{F.~Berghaus}\UBC
% ...
\collaboration{The K2K Collaboration}\noaffiliation


\begin{abstract}
    In this paper we report on the measurement of the rate of
    inclusive $\pi^0$ production induced by charged-current neutrino
    interactions in a C$_8$H$_8$ target at a mean energy of 1.3 GeV in the K2K near detector.
    Out of a sample of 11,606 charged current neutrino interactions,
    we select 479 $\pi^0$ events with two reconstructed photons.
    We find that the cross
    section for the inclusive $\pi^0$ production relative to the
    charged-current quasi-elastic cross section is
    $$\frac{\sigma_{CC\pi^0}}{\sigma_{CCQE}}=0.426\pm0.032(stat.)\pm0.035(syst.)$$
    The energy dependent cross section ratio is also measured.
    The results are consistent with previous experiments for exclusive
    channels on different targets.
\end{abstract}
\maketitle
\section{\label{sec:introduction}Introduction}
After the observation of solar
neutrino and atmospheric neutrino oscillations
\cite{homestake, Fukuda:1996sz, gallex, sage, Fukuda:1998fd, sno, Fukuda:1998mi} and their confirmation respectively at
reactors \cite{kamland} and accelerators \cite{Ahn:2006zz}, the
primary aim of current and future neutrino experiments is to
measure the $\theta_{13}$ mixing angle, and to improve accuracy
in the measurement of oscillation parameters. One of the largest limitations of
accelerator-based neutrino experiments comes from the poor
experimental knowledge of neutrino cross sections in the GeV energy
range. Concerning the measurement of $\theta_{13}$
via sub-leading $\nu_{\mu}\to\nu_e$ oscillation searches, one of the main
backgrounds to the $\nu_e$ signal comes from $\nu_{\mu}$ neutral-current
(NC) interactions producing $\pi^0$'s. Experimental input on the
rate of the related charged-current (CC) channel, which is the focus of this
paper, and measurement of the $\pi^0$ production momentum spectrum,
allows better understanding of this background. Concerning the improvements
in the measurement of oscillation parameters, and in particular of the
atmospheric mixing angle $\theta_{23}$ and mass-squared difference
$\Delta m^2_{23}$ via the measurement of the distortion of the
neutrino energy spectrum induced by neutrino oscillations, 
knowledge of the overall yield and interaction
type composition of CC inelastic interactions is crucial.
This is because the reconstruction of neutrino energy in CC interactions via kinematic means
is less accurate in inelastic interactions, compared to quasi-elastic (CCQE) 
interactions.  Charged-current inclusive $\pi^0$ production (CC$\pi^0$)
constitutes a large component of all CC inelastic interactions.
In addition, since uncertainties in the nuclear models play a significant
role in the neutrino-nucleus cross section, it is important to have
measurements on different target materials. 

Although there are several theoretical approaches to model these processes, the experimental
constraints are rather weak. Very little data exists in the few GeV neutrino
energy range. Experimental measurements of neutral pion production via
CC interactions of few-GeV neutrinos on deuterium have been
collected in the past for single pion \cite{Barish:1978pj,Radecky:1982fn,Kitagaki:1986ct,Kitagaki:1990vs}
and two pion \cite{Day} final states. At higher energies,
CC single $\pi^0$ production cross sections have been measured on
deuterium \cite{Allasia:1990uy} and heavy freon \cite{Grabosch:1988gw} targets.

In this paper we present the measurement of
inclusive CC neutrino interactions with a $\pi^0$ in the final
state made with the K2K SciBar/EC detector system. 
The measurement presented here is the first result
on a carbon target in the few-GeV neutrino energy range, and improves 
the precision of previous results on
different targets. First, we obtain the cross section for this process 
with respect to the cross section for both CC QE and inelastic interactions.
We quote our result as a cross section ratio rather than as
an absolute cross section in order to reduce the impact of large
uncertainties in the estimation of the K2K neutrino flux affecting the
SciBar detector. Second, by reconstructing the neutrino energy of CC interactions
resulting in inclusive $\pi^0$ production, we present the energy
dependence of this cross section ratio. Third, by using previous K2K 
experimental input on CC single pion production \cite{analisa}, we interpret our result
as a measurement of the CC deep inelastic cross section, relative to the
CC QE cross section. Fourth, we present relevant $\pi^0$ production kinematic
distributions of our CC$\pi^0$ candidate events.

The paper is organized as follows: Section \ref{sec:experiment} describes the
experimental setup: the neutrino beam and the neutrino near detector at KEK;
Section \ref{sec:simulation} describes the simulation of the experiment, focusing on the neutrino
interaction simulation. In Section \ref{sec:CCanalysis} we discuss 
the ingredients of our main cross section analysis, describing the experimental signature,
the CC event selection, the photon selection, the $\pi^0$ mass
and the neutrino energy reconstruction; Section
\ref{sec:LikelihoodFit} describes our likelihood fit method; Section \ref{sec:systematic} describes the
systematic uncertainties affecting our measurement; Section \ref{sec:results} presents
the energy-dependent and energy independent cross section results, and the comparison with
the neutrino interaction simulation and with existing results are
given in Section \ref{sec:Comparisonwithotherexperiment}.
Conclusions are given in Section \ref{sec:Conclusion}.

\section{\label{sec:experiment}Experimental setup}
\subsection{Neutrino Beam}
\label{subsec:experiment_beam}

The KEK to Kamioka (K2K) experiment
\cite{Ahn:2006zz,Aliu:2004sq,Ahn:2002up,Ahn:2001cq} is a long
baseline neutrino oscillation experiment in which a beam of muon
neutrinos created at KEK is detected 250 km away in the
Super-Kamiokande detector, located in Kamioka, Japan. To produce the
neutrino beam, protons are accelerated by the KEK proton synchrotron
to a kinetic energy of 12 GeV and then extracted every 2.2 s in a
single turn to the neutrino beam line.  The duration of an
extraction, or spill, is 1.1 $\mu$s and each spill contains 9
bunches of protons at a 125 ns time interval. The
protons are steered to the neutrino beam line to strike an aluminum
target, producing secondary particles. Two toroidal magnetic horns
focus the positively charged particles, mainly $\pi^{+}$'s, in the
forward direction.  The focused positive pions are allowed to decay
into a 200 m long tunnel where they produce neutrinos via $\pi^{+}
\rightarrow \mu^{+}\nu_{\mu}$. A beam absorber made of iron,
concrete, and soil is located at the end of the decay volume to stop
all particles except neutrinos. 
%
%
The direction and intensity of the neutrino beam are checked
spill-by-spill by monitoring the muons produced by pion decay.
The energy spectrum of the neutrino beam is checked by
occasionally monitoring the pions focused by the horn magnets

Over the duration of the K2K experiment, a
total of $9.2\times10^{19}$
protons were delivered to the target to generate
the neutrino beam. The SciBar and EC detectors took data from October 2003 until
November 2004;
$2.02\times$10$^{19}$
protons on target were
accumulated during this time. 

\begin{figure}[ht]
\begin{center}
  \includegraphics[width=0.45\textwidth]{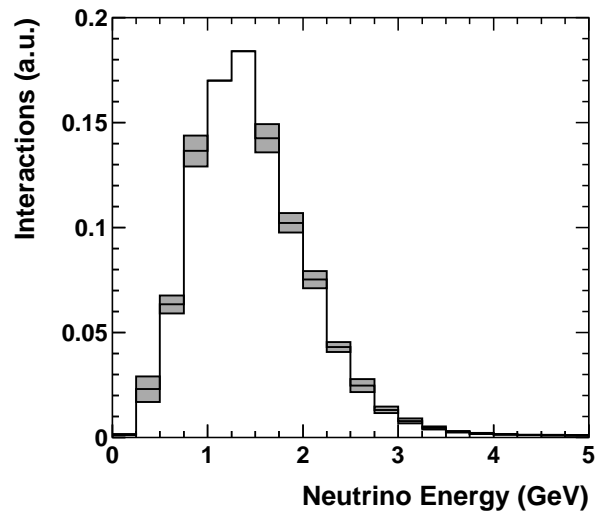}
\caption{The energy spectrum for all muon neutrino interactions
in the SciBar fiducial volume. The gray boxes correspond to the shape systematic uncertainty.} \label{fig:spectrum}
\end{center}
\end{figure}

A Monte Carlo (MC) simulation is used to predict the properties of the neutrino
beam.  According to the simulation, the beam at the near detector is
about 97.3\% pure $\nu_{\mu}$ with a mean energy of 1.3 GeV.
A fit of data of neutrino interactions in all the near detectors
is used to fine-tune the simulated neutrino energy spectrum \cite{Ahn:2006zz}.
Figure \ref{fig:spectrum} shows the energy spectrum for all muon neutrino
interactions in the fiducial volume of the SciBar detector.

\subsection{Neutrino Detectors at KEK}
\label{subsec:experiment_neardet}

The near detector system is located 300 m downstream of
the proton target. The purpose of the near detector is to measure
the direction, flux and energy spectrum of neutrinos at KEK before
oscillation. The near detector is also used for measurements of
neutrino cross sections.

\begin{figure}[ht!]
\begin{center}
  \includegraphics[width=0.45\textwidth]{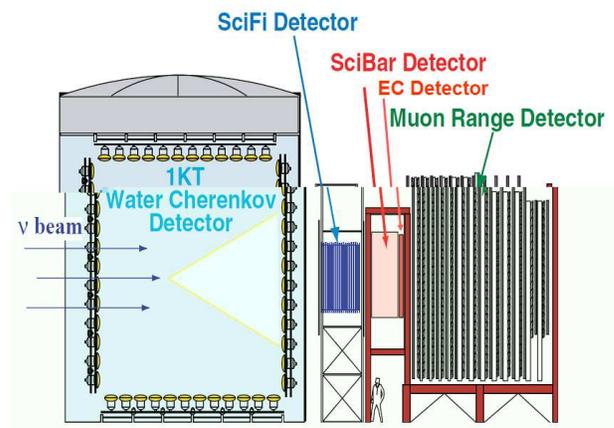}
\caption{Schematic view of the near neutrino detector.} \label{nd}
\end{center}
\end{figure}

A schematic view of the near detector is shown in Fig.~\ref{nd}. The
near detector consists of a one kiloton water \v{C}erenkov detector
(1KT) \cite{Nakayama:2004dp}, a scintillating-fiber/water target
tracker (SciFi) \cite{Suzuki:2000nj}, a fully active
scintillator-bar tracker (SciBar) complemented by a lead and fibers
electron catcher (EC) and a muon range detector (MRD). In this
section we describe the SciBar, EC and MRD since data taken from
these detectors are used in the present analysis. A full description
of the K2K near detectors can be found in \cite{Ahn:2006zz}.

\subsubsection{Scintillator Bar detector (SciBar)}

The SciBar detector acts as a fully active
neutrino target and its primary role is to reconstruct the neutrino
interaction vertex and detect the final state charged particles.

\begin{figure}[htp!]
  \begin{center}
    \includegraphics[width=0.45\textwidth]{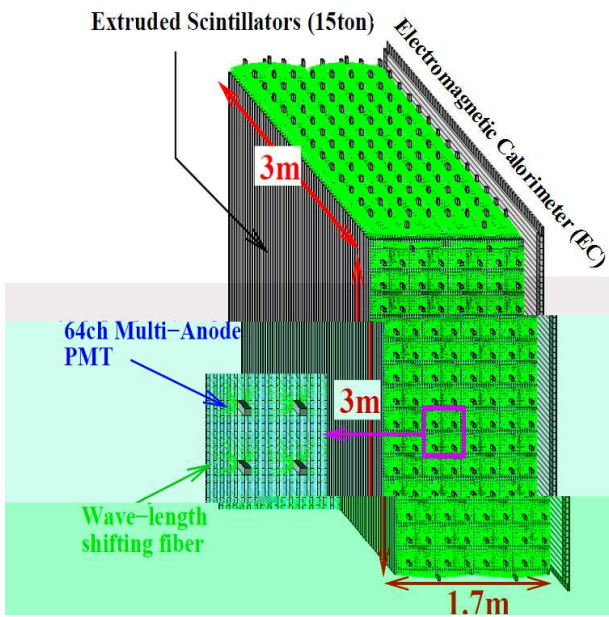}
  \end{center}
  \caption{\label{fig:scibar} Diagram of SciBar and of the Electron Catcher (EC).}
\end{figure}

SciBar \cite{Nitta:2004nt,Yamamoto:2005cy} consists of 14,848 extruded scintillator bars
of 1.3 $\times$ 2.5 $\times$ 300 cm$^3$. Groups of 116 bars
are arranged horizontally or vertically to make one plane.
The planes are arranged in 64 layers orthogonal to the beam, each
consisting of one horizontal and one vertical plane.
The total volume is 3 m $\times$ 3 m $\times$ 1.7 m, for a total mass of $\sim$15 tons. Figure \ref{fig:scibar}
shows a diagram of the SciBar detector.

The extruded scintillator bars are produced by FNAL
\cite{Pla-Dalmau:2001en}. The bars are made of polystyrene
(C$_8$H$_8$), PPO (1\%), and POPOP (0.03\%). Each bar is 1.3 cm
$\times$ 2.5 cm $\times$ 300 cm and has a 0.25 mm thick reflective
coating made of TiO$_2$. The peak of the emission spectrum for the
scintillator is at 420 nm. A 1.5 mm diameter wavelength shifting
(WLS) fiber (Kuraray Y11(200)MS) is inserted in a 1.8 mm hole in
each bar to guide the scintillation light to multi-anode
photomultiplier tubes (MAPMTs). The average attenuation length of
the WLS fibers is approximately 350 cm. 
The absorption peak for the fibers is at 430 nm (matching the
emission peak for the scintillator), and the emission peak is at 476
nm. 
The scintillation light produced is detected by Hamamatsu 
H8804 MAPMTs. Each MAPMT has 64 channels arranged in an 8$\times$8 array. Each
pixel is 2 mm $\times$ 2 mm. The cathode material is Bialkali, with
a quantum efficiency of 21\% at a wavelength of 390 nm. The cathode
is sensitive to wavelengths between 300 and 650 nm. A typical
channel gain is 6$\times$10$^5$ at a supply voltage of 800-900 V. The
basic properties such as gain and linearity are measured for each
channel before installation. The non-linearity of the output signal
vs. input charge is 5\% at 200 photo electrons (p.e.) at a gain of
5$\times$10$^5$. Crosstalk in the MAPMT is approximately 3\% in
neighboring channels. Groups of 64 fibers are bundled together and
glued to an attachment to be precisely aligned with the pixels of the MAPMT.
SciBar's readout system \cite{Yoshida:2004mh} consists of
a front-end electronics board (FEB) attached to each MAPMT and a
back-end VME module. The front-end electronics uses VA/TA ASICs. The
VA is a 32-channel pre-amplifier chip with a shaper and multiplexer.
The TA provides timing information by taking the ``OR'' of 32
channels. Each FEB uses two VA/TA packages to read 64 analog signals
and two timing signals for each MAPMT. Each back-end VME board
controls the readout of eight FEBs. Flash ADCs are used to digitize
the charge information, and TDCs are used to process the timing
information. The pedestal width is approximately 0.3 p.e., and the
timing resolution is 1.3 ns. In order to monitor and correct for
gain drift during operation, SciBar is equipped with a gain
calibration system using LEDs \cite{Hasegawa-thesis}. The system
shows that the gain is stable within 5\% for the entire period of
operation. Cosmic ray data collected between beam spills are used to
calibrate the light yield of each channel. The average light yield
per bar is approximately 20 p.e. for a minimum ionizing particle.
The light yield is stable within 1\% for the whole period of
operation. Pedestal, LED, and cosmic-ray data are taken
simultaneously with beam data.
A crosstalk correction is applied to
both data and MC before event reconstruction \cite{analisa}. After
the crosstalk correction, scintillator strips with a pulse height larger 
than two p.e.(corresponding to about 0.2 MeV) are selected for tracking.
Charged particles are reconstructed by looking for track projections
in each of the two-dimensional views ($x$-$z$ and $y$-$z$) using a
cellular automaton algorithm \cite{Glazov:1993ur}. Three-dimensional
tracks are reconstructed by matching the $z$-edges and timing
information of the 2D tracks. Reconstructed tracks are required to
have hits in at least 3 consecutive layers. The minimum length of a
reconstructible track is, therefore, 8 cm, which corresponds to a
momentum threshold of 450 MeV/$c$ for protons. The reconstruction
efficiency for an isolated track longer than 10 cm is 99\%. The
efficiency is lower for multiple track events due to overlapping of
tracks in one or both views.
\subsubsection{Electron Catcher (EC)}
The EC detector is an electromagnetic calorimeter installed just
downstream of SciBar as shown in Fig.~\ref{fig:scibar}. The main
purpose of the EC is the longitudinal containment of the
electromagnetic showers since the whole SciBar corresponds to only 4
radiation lengths. The EC provides 11 radiation lengths and has 85\%
energy containment at 3 GeV. The EC consists of one plane of 30
horizontal modules and one plane of 32 vertical modules. The two
planes have a cross sectional area of 2.7 m $\times$ 2.6 m and 2.6 m
$\times$ 2.5 m, respectively. The modules were originally made for
the CHORUS neutrino experiment at CERN \cite{Buontempo:1994yp}. Each
module is a sandwich of lead and scintillating fibers, built by
piling up extruded sheets of grooved lead with scintillating fibers
positioned in the grooves. A module consists of a stack of 21 lead
sheets, 2650 mm long, and 740 fibers of 1 mm diameter and 3050 mm
long. The groove diameter is 1.1 mm and the sheet thickness is 1.9
mm. The sheets material is 99\% lead with 1\% antimony content to
improve its mechanical properties. The stack is kept together by a
welded steel case. An overall thickness non-uniformity of less than
2\% was achieved through the extrusion process.At both ends fibers
are bundled in two independent groups, defining two different
readout cells of about $42\times42$ mm$^2$ transverse cross section.
The fibers are manufactured by Kuraray (type SCSF81) and consist of
a polystyrene core surrounded by a 30 $\mu$m thick acrylic cladding,
with an emission maximum in the blue, around 420 nm. To improve the
light collection uniformity, an acrylic black paint is applied on the
surface of the last 5 cm of fibers on each side. This has the effect
of reducing the light coming from the cladding, which has a smaller
attenuation length. In addition, in order to select the spectral
component with a larger attenuation length, a yellow filter (Kodak
Wratten No.3) is used. The attenuation length was measured to be
($462\pm53$) cm when the modules were built \cite{Buontempo:1994yp}
and was recently measured to be ($400\pm12$) cm. At both ends of
the readout cell, fibers are grouped into two bundles of
hexagonal cross-section (22.2 mm apex to apex) and are coupled to a
Plexiglas light guide, also with hexagonal cross section (24 mm apex
to apex). The hexagonal shape and the length of the light guide were
chosen to reduce disuniformities in the mixing of the light coming
out of the individual fibers \cite{acosta}.
The light guides are
coupled to 1-1/8 inch diameter photomultipliers, type R1355/SM from
Hamamatsu, with a special green extended photocathode, of 25 mm
effective diameter. The cathode material is Bialkali, with a quantum
efficiency of 27\% in the wavelength range 350-450 nm. The cathode
is sensitive to wavelengths from 300 nm to 650 nm. A typical current
amplification is $2.1\times10^6$ at the supply voltage of 1600 V. The
anode dark current is 10 nA. The PMT gain of each channel was
measured before installation. The non-linearity of the output
signal vs input charge is 2\% at 60 mA (corresponding to 600 photo
electrons) at a gain of $2\times10^6$.  The
PMT produces a differential signal using the outputs of the cathode
and the last dynode. Signals are read via multipolar differential
screened cables, 100 m long.
The readout system consists of 8 QDC VME (CAEN V792) with 32-channel
12-bit ADC. Impedance matching cards (CAEN A992 custom
modified) are used to convert the differential signals into single
ended signals and to decouple the PMT's and the QDC grounds.
Cosmic rays measured during normal data taking in between the neutrino
spills are used to calibrate the detector and to monitor the gain
stability. After the calibration the spread in the individual
channel response was stable within 1\%. The pedestal width
is approximately 0.7 photo electrons and the energy resolution was
measured in a test beam as $14\%/\sqrt{E\textnormal{(GeV)}}$. The
energy deposited is reconstructed by searching for clusters of
nearby hits above threshold.  In this analysis clusters are
reconstructed searching for hits with more than 20 MeV in the
vertical plane and 10 MeV in the horizontal plane. Hits in the
nearest counters are iteratively added to the cluster if their
energy is greater than 10 MeV(5 MeV) for the vertical(horizontal)
plane. The cluster position is the energy weighted average of the
positions of the counters belonging to the cluster.
\subsubsection{Muon range detector (MRD)} \label{sec:MRD}
The MRD \cite{Ishii:2001sj} is the most downstream detector. It
consists of 12 layers of iron between 13 layers of vertical and
horizontal drift-tubes. Each layer is approximately 7.6 m $\times$
7.6 m. To have good energy resolution over the entire energy spectrum,
the four upstream iron layers are each 10 cm thick, while the other
eight planes are 20 cm thick.  The total iron thickness of 2 m
covers muon energies up to $2.8$ GeV, which corresponds to 95\% of
all the muons produced by neutrino interactions in K2K. There are
6,632 aluminum drift tubes filled with P10 gas (Ar:CH$_4$ =
90\%10\%). The total mass of the iron is 864 tons, and the mass of
the drift tubes is 51 tons.
The MRD is used to monitor the stability of the neutrino
beam direction, profile and spectrum by measuring the energy, angle
and production point of muons produced by CC
neutrino interactions in the iron target. The MRD is also used to
identify muons produced in the upstream detectors.
The energy and angle of the muon can be measured by the combination
of the MRD and the other fine-grained detectors. It is necessary to
measure the muon energy and direction in order to reconstruct the
energy of the incident neutrino for CC events. The MRD tracking
efficiency is 66\%, 95\%, and 97.5\% for tracks that traverse one,
two and three iron layers, respectively; for longer tracks, the
efficiency approaches 99\%.  A track that hits less than three
layers of MRD is called "one layer hit" (MRD1L), while a track that
hits more than three iron layers will be reconstructed as a 2D track
in xz or yz planes. The 2D-tracks pair which has the longest overlap
is taken as a 3D-track (MRD3D). The range of a track is estimated
using the path length of the reconstructed track in iron. The muon
energy is calculated by the range of the track. The uncertainty in
the muon energy due to differences among various calculations of the
relationship between muon energy and range is 1.7\%. The uncertainty
in the weight of the iron is 1\%. Thus, the systematic error in the
MRD energy scale is quoted as the sum of these uncertainties, 2.7\%.
The energy resolution is estimated by Monte Carlo simulation to be
0.12 GeV for forward-going muons. The angular resolution is about 5
degrees.


\section{\label{sec:simulation}Simulation}
\subsection{Neutrino Interactions}
\label{subsec:simulation_interactions}

The neutrino interaction simulation plays an important role
for estimating the event yields, and the
topological and kinematical properties, for CC
neutrino interactions in SciBar producing neutral pions, as well
as for background processes. We use the
NEUT program library to simulate neutrino
interactions with protons and carbon nuclei within the SciBar detector
material. NEUT \cite{Hayato:2002sd} simulates neutrino interactions
over a wide energy range, from $\sim$100 MeV up to TeV neutrino energies, and
on different nuclear targets. 

In the simulation program, the following CC and NC neutrino
interactions are considered: QE scattering ($\nu\ N \rightarrow l\
N'$), single pion production ($\nu\ N \rightarrow l\ N'\ m$),
coherent $\pi$ production ($\nu\ ^{12}{\rm C} \rightarrow\ l\ \pi\
^{12}{\rm C}$), and deep inelastic scattering (DIS, $\nu N\
\rightarrow l\ N'\ hadrons$). In these reactions, $N$ and $N'$ are
the nucleons (proton or neutron), $l$ is the lepton (either a
charged lepton or a neutrino), and $m$ is a meson. If the neutrino
interaction occurs in a carbon nucleus, the interactions of the
generated particles with the remaining nucleons of the nucleus are
also simulated.

\begin{figure}[htbp]
\begin{center}
\includegraphics[width=0.45\textwidth]{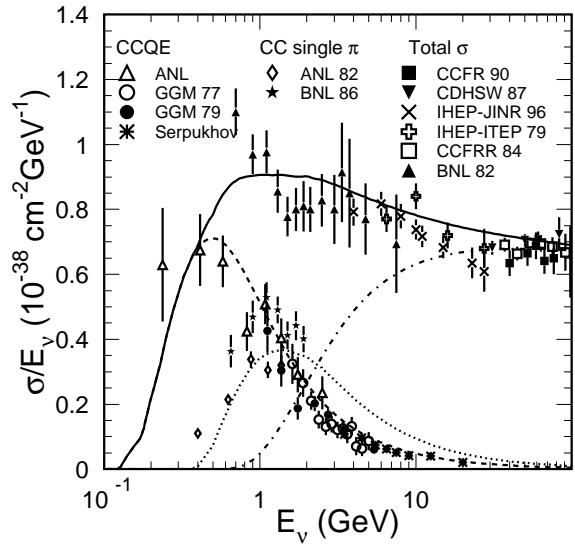}
\caption{Charged-current total cross section divided by the neutrino
energy $E_\nu$ for neutrino-nucleon charged-current interactions\cite{Hayato:2002sd}.
The solid line shows the calculated total cross section. The dashed,
dotted and dash-dotted lines show the calculated quasi-elastic,
single pion and deep inelastic scattering, respectively. 
The data points 
are taken from the following experiments: 
($\triangle$) ANL\protect\cite{Barish:1977qk},
($\bigcirc$) GGM77\protect\cite{Bonetti:1977cs},
($\bullet$) GGM79 (a)\protect\cite{Ciampolillo:1979wp}, (b)\protect\cite{Armenise:1979zg},
($\ast$) Serpukhov\protect\cite{Belikov:1985kg},
($\Diamond$) ANL82\protect\cite{Radecky:1982fn},
($\star$) BNL86\protect\cite{Kitagaki:1986ct},
($\blacksquare$) CCFR90\protect\cite{Auchincloss:1990tu},
($\blacktriangledown$) CDHSW87\protect\cite{Berge:1987zw}, 
($\times$) IHEP-JINR96\protect\cite{Anikeev:1996dj},
($+$) IHEP-ITEP79\protect\cite{Mukhin:1979bd},
($\Box$) CCFRR84\protect\cite{MacFarlane:1984ax}, and
($\blacktriangle$) BNL82\protect\cite{Baker:1982ty}.
}
\label{plot_tot}
\end{center}
\end{figure}

The total charged-current cross section predicted by NEUT, together
with the QE scattering, single pion production and deep inelastic
scattering contributions, are shown in Fig.~\ref{plot_tot},
overlaid with data from several experiments.

Given the K2K beam neutrino energy spectrum, Table~\ref{tab:nuint}
shows the fraction of interactions in SciBar that are expected to be
QE, single pion, etc. according to the simulation.

\begin{table}[htbp]
\begin{tabular}{l|r}
\hline
\hline
\textbf{~~Interaction type}&\textbf{Percent of Total}\\
\hline
\hline
\textbf{Charged-current (CC)}&\textbf{71.8\%}\\
~~~~~~$\nu_{\mu} n \rightarrow \mu^{-} p$ (QE)&~~~~32.2\%\\
~~~~~~$\nu_{\mu} p \rightarrow \mu^{-} p \pi^{+}$&~~~~18.0\%\\
~~~~~~$\nu_{\mu} n \rightarrow \mu^{-} n \pi^{+}$&~~~~6.2\%\\
~~~~~~$\nu_{\mu} n \rightarrow \mu^{-} p \pi^{0}$&~~~~5.0\%\\
~~~~~~DIS &~~~~6.8\%\\
~~~~~~others (K and $\eta$) &~~~~3.6\%\\
\textbf{Neutral-current (NC)}&\textbf{28.2\%}\\
\hline
\hline
\end{tabular}
\caption{\label{tab:nuint}Expected neutrino interactions in SciBar.}
\end{table}

\subsubsection{\label{subsubsec:simulation_interactions_ccpi0}
$\pi^0$-producing charged-current neutrino interactions}
For the simulation of CC neutrino interactions resulting
in inclusive $\pi^0$ production, we adopt distinct models, depending
on the invariant mass $W$ of the hadronic system in the final state,
and on the pion multiplicity. A summary of the models used to simulate
the cross section and the final state kinematics is given in
Table~\ref{tab:nuint_signaldef},
while more details are given in the text below.

\begin{table}[htbp]
\begin{tabular}{|c|c|c|c|}
\hline
\hline
\textbf{$W$}&\textbf{Process}&
\textbf{Cross}&\textbf{Final State}\\
\textbf{(GeV/c$^2$)}&\textbf{}&
\textbf{Section}&\textbf{Kinematics}\\
\hline
\hline
 $<$2.0 & $\nu_{\mu} n \rightarrow \mu^{-} p \pi^{0}$ &
\cite{Rein:1981wg} & \cite{Rein:1987cb}, isotropic \\ \hline
1.3-2.0 & $\nu_{\mu} N \rightarrow \mu^{-} N' \pi^{0}\pi$ & 
\cite{Gluck:1994uf,Bodek:2002vp} & 
\cite{Nakahata:1986zp} \\ \hline
$>$2.0 & $\nu_{\mu}N\to\mu^{-}N'\pi^{0}X$ & 
\cite{Gluck:1994uf,Bodek:2002vp} & 
\cite{Sjostrand:1994yb} \\
\hline
\hline
\end{tabular}
\caption{\label{tab:nuint_signaldef}Models used to simulate the cross section
and final state kinematics for CC inclusive $\pi^0$
production. In the table, $W$ stands for the invariant mass of the final
state hadronic system, $N$ and $N'$ for nucleons, $\pi$ for
at least one charged or neutral pion, and $X$ for any meson (including none).
See text for details.}
\end{table}

For $W<2$ GeV, and production of single $\pi^0$'s and no other
pions (charged or neutral), we use the resonance-mediated
Rein-Sehgal model~\cite{Rein:1981wg}. In this model, the interaction
simulation is performed via a two-step process. First, the neutrino-induced
excitation of the baryon resonance $N^*$ is modeled:
\begin{equation*}
  \nu_{\mu} + n \rightarrow \mu^- + N^*,
\end{equation*}
which is then followed by the resonance decay to a pion-nucleon final
state:
\begin{equation*}
  N^* \rightarrow \pi^0 + p.
\end{equation*}

The same $\pi^0$-$p$ final state
can be fed by several resonances. All baryon resonances with $W<$ 2
GeV/c$^2$ are taken into account with their corresponding resonance width
and including possible interferences among them.
Single $K$ and $\eta$ productions are
simulated by using the same framework as for the dominant
single $\pi$ production processes. The model
contains a phenomenological parameter (the single pion axial vector mass,
$M_{A}$), that must be determined experimentally. As the value of
$M_{A}$ increases, interactions with higher $Q^2$ values (and
therefore larger scattering angles) are enhanced. The $M_{A}$
parameter in our resonance-mediated Rein-Sehgal model is set to
1.1~GeV/c$^2$. To determine the final
state kinematics in the decay of
the dominant resonance $P_{33}$(1232), Rein's method \cite{Rein:1987cb} is used to
generate the pion angular
distribution in the resonance rest frame. For the other resonances,
the directional distribution of the generated pion is set to be
isotropic in the resonance rest frame. The pion angular distribution
for the related $\nu p \rightarrow \mu^- p \pi^+$
mode has been measured~\cite{Kitagaki:1986ct},
and the results agree well with our
model. To describe nucleons bound in carbon nuclei, nucleons are
treated as quasi-free particles in motion using
a relativistic Fermi gas model~\cite{Smith:1972xh,Moniz:1971mt}
with 225~MeV/$c$ Fermi surface momentum, and assuming
a nuclear binding energy of 27~MeV. 
The Pauli blocking effect in the decay of
the baryon resonance is taken into account by requiring that the
momentum of the nucleon should be larger than the Fermi surface
momentum. In addition, pion-less decay for the dominant $\Delta$
resonance
($\Delta N\to NN$) is considered. In this case, which is
expected to occur with a 20\% probability,
no pion is present in
the final state; only a lepton and a nucleon are
emitted~\cite{Singh:1998}.

For the production of $\pi^0$'s in association
with other pions (charged or neutral), the deep inelastic scattering
cross section formalism
combined with GRV94 parton distribution
functions \cite{Gluck:1994uf} is used. Additionally, we have included
the corrections in the small $Q^2$ region developed by Bodek and
Yang~\cite{Bodek:2002vp}.
For the simulation of DIS final state kinematics in which the
hadronic invariant mass, $W$, is larger than $2$ GeV/$c^2$, we use the
PYTHIA/JETSET library~\cite{Sjostrand:1994yb}. For $W<2$ GeV/$c^2$, we
use a custom made program \cite{Nakahata:1986zp}, based
on data-driven average pion multiplicities and KNO scaling.
In the latter case,
the multiplicity of pions is required to be larger than one because
single pion production is already taken into account by the
resonance-mediated single pion production.

\subsubsection{\label{subsubsec:simulation_interactions_other}
Other neutrino interactions}
Resonance-mediated single pion production and deep inelastic
scattering CC processes that do not result in the production of
$\pi^0$'s, but possibly in the production of other mesons such as $\pi^{\pm}$,
are also  simulated according to the models described in
Sec.~\ref{subsubsec:simulation_interactions_ccpi0}. The same
models are used as well to simulate the corresponding NC
channels.

The formalism of CC and NC QE
scattering off free nucleons used in
the simulation is described by
Llewellyn-Smith~\cite{LlewellynSmith:1972zm}. There is only one
parameter in the model to be
determined experimentally, the QE axial vector mass, M$_A$.
As for single pion production via baryon resonances, M$_A$ is set to
1.1 GeV/$c^2$ in our simulation, based on near detector data
\cite{Ahn:2002up}.

Coherent single pion production, that is the
interaction between a neutrino and the entire carbon nucleus
resulting in the production of single pions and no nuclear
break-up, is
simulated using the formalism developed by Rein and Sehgal
\cite{Rein:1983pf}. The coherent pion production axial vector mass
is set to 1.0 GeV/c$^2$ in our model.
Only neutral-current coherent pion production interactions are
considered, because the cross section of the CC coherent
pion production was found to be very small at K2K beam
energies~\cite{Hasegawa:2005td}.

\subsubsection{Intra-nuclear hadronic interactions}
The intra-nuclear interactions of the mesons and nucleons produced
in neutrino interactions with carbon nuclei are
also important for this analysis.  
Due to the propagation in the nuclear matter of the target nucleus, the final
state particles observed differ from 
the one produced at the weak interaction vertex.
Particles absorption or production as well as changes in the direction 
or momentum affect the event classification.
For example, $\pi^0$'s produced at the weak interaction vertex
can be absorbed via intra-nuclear interactions within the target nucleus,
therefore escaping direct detection. Likewise, intra-nuclear interactions
can result in $\pi^0$ production within the target nucleus, even in the absence
of $\pi^0$'s at the weak interaction vertex.
Therefore, the interactions of pions, kaons, etas and nucleons are
also taken into account. The meson and nucleon interactions are treated using
a cascade model, and each of the particles is traced in the nucleus
until escaping from it.

In our simulation, the following intra-nuclear pion interactions
are considered: inelastic scattering, charge exchange and
absorption. The actual procedure to simulate these interactions is
the following: first, the generated position of the pion in the nucleus
is set according to the Woods-Saxon nucleon density
distribution~\cite{Woods:1954}. Then, the interaction mode is
determined by using the calculated mean free path of each
interaction.  To calculate these mean free paths, we adopt the model
described by Salcedo et al.~\cite{Salcedo:1988md}. The calculated
mean free paths depend not only on the momentum of the pion, but also
on the position of the pion in the nucleus. If inelastic scattering or
charge exchange occurs, the direction and momentum of the pion is
determined by using the results of a phase shift analysis obtained
from $\pi-N$ scattering experiments~\cite{Rowe:1978fb}.  When
calculating the pion scattering amplitude, the Pauli blocking effect
is also taken into account by requiring the nucleon momentum after
the interaction to be larger than the Fermi surface momentum at the
interaction point. This pion interaction simulation is tested by
comparison with data, including $\pi
^{12}$C scattering and pion photo-production
($\gamma + ^{12}$C $\rightarrow \pi^- + X$) data.

Re-interactions of the
nucleons (protons, neutrons) produced in the neutrino
interaction are also important. Each nucleon-nucleon interaction modifies the
nucleon momentum and direction, possibly causing the number of visible
nucleons to be mis-predicted if not properly
modeled~\cite{Walter:2002sa}. Elastic scattering, single and two-pion
production, are considered.

Our simulation predicts that in 26\% of SciBar CC interactions with
$\pi^0$ production at the primary neutrino-nucleon electroweak
vertex, the neutral pion(s) does not escape the target nucleus.
On the other hand, 15\% of the events with $\pi^0$ emerging from
the target nucleus are produced in nuclear interactions.

\subsection{Detector Response}
\label{subsec:simulation_detector}

The GEANT3 \cite{Brun:1987ma} package is used to simulate the
detector geometry and the interactions and tracking of particles.
The CALOR program library \cite{Zeitnitz:1994bs} is used to simulate the
interactions of pions with the detector material for pions with
momentum greater than 0.5~GeV/c.  For lower energy pions, a custom
library \cite{Nakahata:1986zp} is used.

The energy loss of a particle in each single SciBar strip and each
individual EC sensitive fiber is simulated. The energy deposition is
converted in the detector response taking into account
the Birk's saturation of the scintillator, the light attenuation
along the fibers, the Poisson fluctuation of the number of photo
electrons, the PMT resolution, and the electronic noise. 
The crosstalk 
in nearby SciBar channels is also taken into account.

In SciBar
the timing of each hit is simulated from the true time of the
corresponding energy deposition, corrected by the travel time of the
light in the WLS fiber and smeared by the timing resolution. 

The MRD simulation includes both ionisation and multiple scattering
in the drift chambers.

The input parameters of the detector simulation are derived from
laboratory measurements and calibration data. The features of the
simulation have been systematically compared and tuned with cosmic
ray and neutrino data.

\section{\label{sec:CCanalysis}Experimental signature}
\subsection{\label{sec:Signaldefinition} Definition of signal and
background}

In this analysis the process we want to measure is defined
inclusively with respect to a $\pi^0$ produced in the target
nucleus of the CC neutrino interaction.  We call signal
an event with one (or more) $\pi^0$ that comes from the neutrino
interaction vertex or from a re-interaction inside the target nucleus. An
event with an $\eta$ decaying into one or more $\pi^0$'s or into
a $\gamma$ pair at the target nucleus is also considered a signal event.
Events are considered background if the final state $\pi^0$ is only produced
due to  
secondary interactions occurring outside of the target nucleus
such as, for instance,
charge exchange of a charged pion,
or $\pi^0$ production in inelastic hadronic interactions. 
Another background category consists of the events selected accidentally
where no $\pi^0$ was produced.

According to this definition, the
CC$\pi^0$ fraction predicted by the neutrino Monte Carlo,
integrated over the K2K energy spectrum, is 13.9\% of the total
number of neutrino CC interactions. The composition of
the signal is the following:

\begin{itemize}
\item 6.5\% is resonant production: 5\% with a $\pi^0$
produced in the resonance decay and 1.5\% with $\pi^0$
produced in a nuclear re-interaction in the target nucleus;
\item 6.6\% is non resonant production, mainly DIS: 
6.0\% with one or more $\pi^0$'s produced at the
neutrino interaction vertex and 0.6\% in nuclear re-interactions
in the target nucleus;
\item 0.8\% comes from nuclear re-interactions, mostly CCQE, where
a $\pi^0$ is produced in the re-interactions in the target nucleus.
\end{itemize}

The fraction of signal events with more than one $\pi^0$ is 43\%.

The final state topology of the CC
inclusive events is characterized by one muon and at least two
electromagnetic showers, plus possibly other particles coming from
the neutrino interaction vertex. If the photon converts in SciBar,
the hit patterns of the low energy electromagnetic showers are
reconstructed by the SciBar tracking algorithm and the direction of
the photon is given by the corresponding track. The SciBar
conversion length is about 40-50 cm (SciBar in fact corresponds to 4
$X_0$). If the photon converts in the EC, the energy is reconstructed
by the EC cluster algorithm and the position of the photon
conversion is the energy-weighted average hit position in the
cluster. Therefore the experimental signature is given by one track
originating in SciBar and reaching MRD, at least two photons
reconstructed either as SciBar tracks disconnected and pointing to
the neutrino interaction vertex or as clusters in the EC. To
isolate a sample of events that satisfy the topology described
above, we first select a clean sample of CC events, characterized by
a SciBar track matched with an MRD track. In
Sec.~\ref{subsec:CCnormalization} we describe the selection criteria to
isolate the CC inclusive sample which is used for normalization. In the same
section, we further classify CC events into sub-samples of
varying CCQE purities, which are used to quote
the CC$\pi^0$ production cross section relative to either
the CCQE or the inelastic cross section. Out of the CC
inclusive sample, we require further cuts to select photons and to separate the
CC$\pi^0$ sample from other topologies. This is described in
Secs.~\ref{subsec:Gamma selection} and \ref{sec:P0reconstruction}.
Section~\ref{sec:enureconstruction} describes the reconstruction of the
incoming neutrino energy.

\subsection{\label{subsec:CCnormalization} CC event selection}

The selection of a CC interaction
requires a muon candidate in the event. A muon
candidate is a reconstructed 3D track in the SciBar fiducial volume
(FV) matching a reconstructed track in MRD. The FV is applied
requiring the upstream edge of the track to be within $\pm$135 cm in
x and y, and -75 $<$ z(cm) $<$ 70 with respect to the center of SciBar. This
corresponds to a 10.9 m$^3$ fiducial volume and 11.6 tons of fiducial
mass. The track is also required to be in time with the neutrino
beam, i.e. within $\pm 50$ ns with respect to the closest neutrino
bunch. The extrapolation of the SciBar track is required to be
matched with a track in MRD. The matching is with a MRD3D track or
with a MRD1L hit, defined in Sec.~\ref{sec:MRD}.

The neutrino
interaction vertex is reconstructed as the upstream edge of the muon
candidate track in SciBar. The resolution in x and y is symmetric
with 0.9 cm RMS. The resolution in z has a 1.6 cm RMS and a small
satellite peak one SciBar layer (2.6 cm) upstream of the true
neutrino vertex, due to crosstalk between MAPMT channels.

We select 11,606 events in
the data and 432,856 in the full MC sample (before normalization),
with an estimated selection
efficiency of 49.5\% and a CC purity of 97.5\%. The main background
comes from the neutral current (NC) multi pion or single pion events
in which a pion gives a signal in the MRD detector. The background
induced from neutrons coming from the beam target
is found to be negligible.

In this analysis we consider four CC sub-samples, shown in
Table~\ref{tab:ccqereduction}, which are characterized by different
fractions of non-QE (nQE) and QE interactions. The first sample
consists of events with a single reconstructed track and has 72.4$\%$
efficiency and 66$\%$
purity for QE events. For the events with two tracks, the direction
of the second track is compared with the expected direction of the
proton in the assumption of a CC QE interaction. If this angle
$\Delta \theta_p$ is smaller than $20^\circ$, the events are
classified as "two tracks quasi-elastic". Events with $\Delta
\theta_p > 20^\circ$ are further divided in two categories,
depending on whether the dE/dx of the second track is consistent
with a pion or with a proton.
\begin{table}[htbn!]
\begin{center}
\begin{tabular}
    {c|c|c|c|c} \hline\hline
    CC sub-sample          & $\epsilon_{QE}$\%   &$\eta_{QE}$\%  &Data & MC$_n$\\ \hline
    1 track                                 &72.4    &66  &6125 & 6080 \\ \hline
    2 tracks QE    &16.9    &76  &1262 & 1307 \\ \hline
    2 tracks nQE $\pi$    &2.3   &12       &1048 & 960 \\ \hline
    2 tracks nQE   p        &7.3                &27       &1453 & 1220 \\
    \hline\hline
\end{tabular}

\caption{Efficiency ($\epsilon_{QE}$) relative to all QE events selected in the CC sample and purity ($\eta_{QE}$) for
QE events, and
number of events selected in MC and in data for the different QE and
nQE samples} \label{tab:ccqereduction}
\end{center}
\end{table}
The Monte Carlo is normalised to data using the first two samples
in Table~\ref{tab:ccqereduction} which have the largest quasi-elastic
contribution. The same normalisation is used in all plots before the fit.
In all plots signal and different background components are stacked.
In order to extract the result
in Sec.~\ref{sec:LikelihoodFit}, we use the four samples
described in Table~\ref{tab:ccqereduction} and we leave the data
to MC normalization free in the fit to properly account for the
correlation between the normalization and the other sources of
systematic error.

\subsection{\label{subsec:Gamma selection} Photon selection}
\begin{figure}[htbp!]
\begin{center}
    \includegraphics[width=0.45\textwidth]{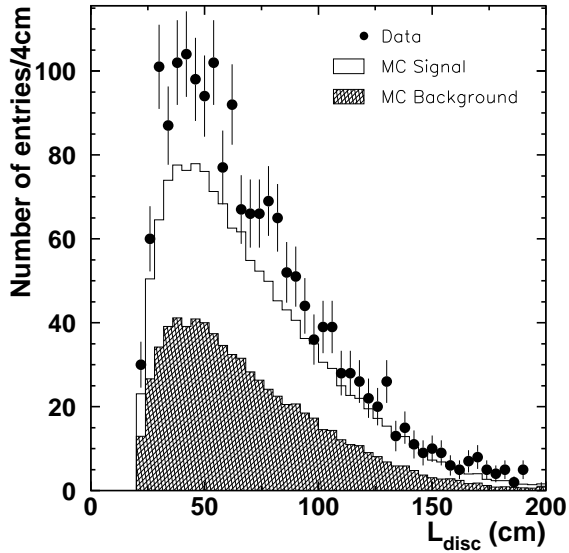}
    \caption{Disconnection: distance L$_{\hbox{disc}}$ between the 
    reconstructed photon candidates
    conversion point and the neutrino interaction vertex.}
    \label{fig:discall}
\end{center}
\end{figure}

All CC selected events are subject to further selection
criteria to tag photons. A photon candidate can be either a
SciBar track or an EC energy cluster. In order to be considered as a
photon candidate a SciBar track should satisfy the following
requirements. First, the timing of the track has to be within 10 ns with
respect to the muon track; second, the track is required not to be
matched with a MRD3D track. Third, the photon conversion point, defined as
the track edge closest to the neutrino interaction vertex, is
required to be within $\pm$145 cm in x and y and $\pm$80 cm in z
with respect to the SciBar center. Fourth, in both
projections the distance between the photon conversion point and the
neutrino vertex has to be larger than 20 cm; and fifth, the track
extrapolation to the Z position of the neutrino vertex should be
within 25 cm from the neutrino vertex. The disconnection from the
vertex of SciBar photon candidates, defined as the 3D distance
between the reconstructed neutrino vertex and photon conversion
point, is shown in Fig.~\ref{fig:discall}.
The shape of the disconnection is consistent with MC. A fit with an
exponential function between 50 cm and 130 cm
gives a value of $\lambda_{\hbox{disc}}=(51.8\pm5.4)$ cm for data, in good
agreement with $\lambda_{\hbox{disc}}=(52.9\pm0.6)$ cm for MC. The result is
consistent with the electromagnetic origin of the selected photon
sample. In fact, according to MC, 82\% of the background events also
contain a genuine photon. 

A fraction of photons converted in SciBar
will have some energy leakage in the EC. Moreover, all the photons
not converted in SciBar and pointing to the EC will convert in the
upstream (vertical) EC plane. We consider only clusters with energy
larger than 50 MeV for the vertical plane and 25 MeV for the
horizontal plane.
The energy of the SciBar photon candidates and the associated EC
vertical and horizontal clusters are added together in order to
reconstruct the photon energy. EC clusters that do not match any of
the reconstructed SciBar tracks are considered isolated. Isolated
vertical clusters are paired to isolated horizontal clusters
according to their energy and they are considered as additional 
photon candidates.

\begin{figure}[htbp!]
\begin{center}
    \includegraphics[width=0.45\textwidth]{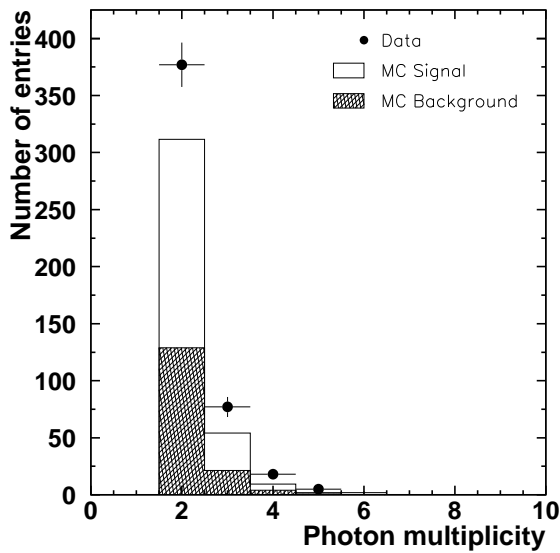}
    \caption{Photon multiplicity of CC$\pi^0$ candidate events}
    \label{fig:gammared}
\end{center}
\end{figure}

Overall, 479 events with at least two photons are reconstructed in data and
380 in MC, with an overall efficiency of 7.6\% and a purity of
59.2\% (MC has been normalized to the data using the normalization
factor described in Sec.~\ref{subsec:CCnormalization}). Figure
\ref{fig:gammared} shows the multiplicity of photon candidates per
event.

The excess of data with respect to the MC is $26\pm6$\%
(statistical error only). 
In 74\% of the candidate events, all photons are converted 
and reconstructed 
in SciBar with possibly an energy leakage in EC. 
In 20\% of the candidate events, one of the photons is converted and
reconstructed in EC, and in 6\% two or more photons are converted 
and reconstructed in EC.
These relative fractions are well reproduced by the MC. It is worth
noting that this strongly supports the hypothesis that the excess of photon candidates is
due to physics and not to detection bias, since SciBar and EC
are completely independent detectors, with different reconstruction
efficiencies and systematics. 

As a crosscheck, we eye-scanned 100
data and 100 MC events. Despite the limited statistics and the
subjectivity in the eye-scan classification criteria, the result is
that the main features of the selected sample are well reproduced by
the MC simulation. In particular, the background coming from the
secondary interactions in data and MC agrees within the statistical
uncertainties of this eye-scan cross-check.

\subsection{\label{sec:P0reconstruction} $\pi^0$ reconstruction}

The $\pi^0$ mass is reconstructed from the energy and the
direction of its two photon decay products:
\begin{eqnarray}
M_{\pi^0}=\sqrt{2 \cdot E_{\gamma_1}\cdot
E_{\gamma_2}\cdot(1-cos(\theta_{\gamma_1,\gamma_2}))}
\end{eqnarray}
where $E_{\gamma_1}$ and $E_{\gamma_2}$ are the reconstructed
energies of the two photons, and $\theta_{\gamma_1,\gamma_2}$ is the
opening angle between them. If the photon converts in SciBar,
the direction is reconstructed using the SciBar 3D reconstructed
track. If the photon converts in EC, we take as photon direction the
direction of the line connecting the reconstructed neutrino vertex
and the center of the EC cluster. For the highest (lowest) energy
photon reconstructed in SciBar, the energy and angular resolutions
(FWHM) are 50 MeV (65 MeV) and 0.15 rad (0.18 rad), respectively. 

As shown in Fig.~\ref{fig:gammared}, 
in 21.4\% of the selected events there are more
than two photon candidates and therefore more than one $\pi^0$ candidate. For
these events the photon pair corresponding to the best $\pi^0$
candidate is selected as the combination which has the reconstructed
$\pi^0$ vertex closest to the neutrino interaction vertex. If
there are one or more EC photon candidates (5.7\% of the total
sample) the best combination is
selected as the photon pair with the reconstructed invariant mass
closest to the $\pi^0$ mass. 

\begin{figure}[htbp!]
  \begin{center}
    \includegraphics[width=0.45\textwidth]{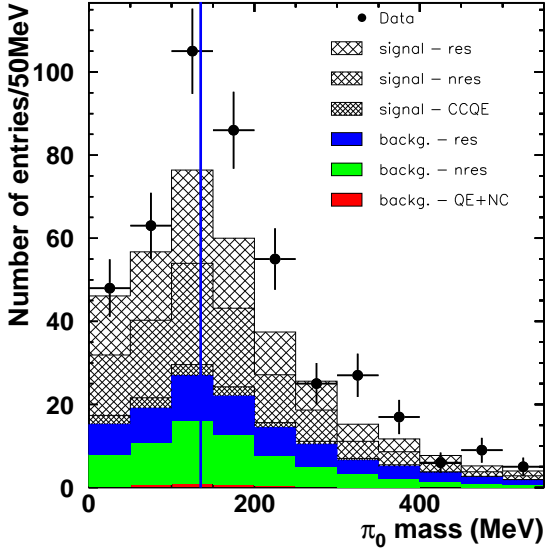}
    \caption{Reconstructed $\pi^0$ mass before fit.}
    \label{fig:p0recbeffit}
  \end{center}
\end{figure}

Figure \ref{fig:p0recbeffit}
shows the reconstructed $\pi^0$ invariant mass for data and
different MC contributions to signal and background. The signal
contribution (according to the definition given in
Sec.~\ref{sec:Signaldefinition}) is divided into $\pi^0$ from resonant and 
non-resonant production and CCQE, and the background is divided into resonant, non resonant production and CCQE plus
neutral current (NC). 
It should be noted that most of the background contains a
$\pi^0$ in the final state, so the shape of the invariant mass
distribution for signal and background is similar.

\subsection{\label{sec:enureconstruction} Neutrino energy reconstruction}

The neutrino energy in a CC interaction can be
reconstructed from the measured muon energy and angle using the
following formula, provided the invariant mass $W$ of the hadronic final state 
is known:

\begin{eqnarray}
E_\nu^{rec}=\frac{(W^2-m_\mu^2)+2E_\mu(M_n-V)-(M_n-V)^2}{2\times(-E_\mu+(M_n-V)+p_\mu
cos(\theta_\mu))} \label{eq:neutrinoenergy}
\end{eqnarray}

\noindent where $V$ is the nuclear potential for carbon which is set to zero, 
and $p_{\mu}$, $E_\mu$ and $\theta_{\mu}$
are the muon momentum, energy and angle. For the QE final state, we have
$W^2=M_p^2$ and the formula used for neutrino energy reconstruction 
in the oscillation analyses. In the present analysis, 98\% of the selected sample is
non-QE, mostly resonant single pion production and DIS,  
and it is characterized by a broad $W$ spectrum.
We found $W=1.483$ GeV the optimal value to
reconstruct the neutrino energy in the MC sample of selected events. We use this value of $W$ 
to reconstruct the neutrino energy in data and Monte Carlo.
The uncertainties on the values assumed for $W$ and for $V$ will be considered 
as a source of systematic errors and evaluated in Sec.~\ref{sec:systematic}.
The resolution turns out to be 
$22\%/\sqrt{E(GeV)}$ for the selected sample. 
The assumption of an average $W$ value is the largest effect in 
the reconstructed neutrino energy resolution. Using the true $W$ value in 
Eq.~\ref{eq:neutrinoenergy} the resolution is $15\%/\sqrt{E(GeV)}$.

\begin{figure}[htbp!]
\begin{center}
    \includegraphics[width=0.45\textwidth]{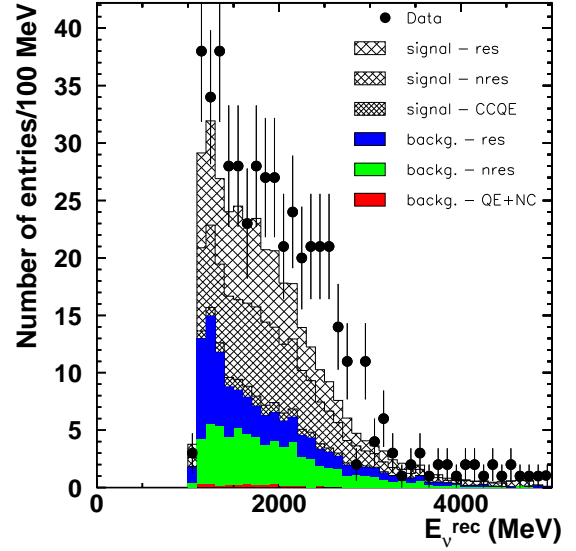}
    \caption{Reconstructed neutrino energy before fit.}
    \label{fig:neutrinoenergy}
    \end{center}
\end{figure}

The reconstructed neutrino energy is shown in
Fig.~\ref{fig:neutrinoenergy} for data and different MC signal 
and background components. The threshold at about 1 GeV is due
to the fixed value assumed for $W$ in Eq.~\ref{eq:neutrinoenergy}.

\section{\label{sec:LikelihoodFit} Likelihood Fit}
From our sample of selected events we measured the ratio
of the inclusive CC$\pi^0$ cross section to the CCQE cross section.
The uncertainties in the absolute neutrino flux cancel out in
the ratio with respect to an independent and relatively well known process like 
the CCQE interaction.

We performed a maximum likelihood fit of the reconstructed neutrino 
energy distribution of the CC$\pi^0$ sample shown in Fig. \ref{fig:neutrinoenergy}. 
At the same time we fit the number of events in the different CC sub-samples 
described in Sec. \ref{subsec:CCnormalization}. 
The likelihood function is defined as

\begin{eqnarray}
L=L_{\pi^0} \cdot L_{CC} =\prod_k P(n_k, \mu_k) \cdot
\prod_s P(n_s, \mu_s) \label{eq:likeli}
\end{eqnarray}

\noindent where $P(n,\mu)$ is the Poisson probability for $n$ observed events
with expectation value $\mu$. The maximum likelihood fit is calculated by minimizing the
log-likelihood function $F=-2Log(L)$ which follows a $\chi^2$ distribution.

The index $s$ labels the 4 CC sub-samples in Table~\ref{tab:ccqereduction}
and the index $k$ labels 50 bins spanning the range 0-5 GeV of 
the reconstructed neutrino energy. 

The expected events $\mu_s$ in each
CC sub-sample (1-track, 2-track QE, 2-track nonQE pion and 2-track nonQE proton)
are defined as:

\begin{eqnarray}
\nonumber \mu_s &=& f_n \{ S_{CC,s}^{QE}+R_{res}S_{CC,s}^{res}+R_{nres}S_{CC,s}^{nres}+\\
&+& B_{CC,s}^{NC}\}
\end{eqnarray}

\noindent where the number of Monte Carlo events contributing to signal ($S_{CC}$) and 
background ($B_{CC}$) are divided into quasi elastic ($QE$), 
resonances production ($res$), non resonant production ($nres$) and 
neutral current processes ($NC$). 
The non resonant production includes all CC non-QE processes
different from resonant production, mainly deep inelastic scattering.  

The parameters $R_{res}$ and $R_{nres}$ are free in the fit in order to 
independently re-weight 
the corresponding Monte Carlo contributions relative to the quasi elastic
process.

All the MC distributions are normalized as described at the end of section~\ref{subsec:CCnormalization}. 
An additional overall normalization parameter $f_n$ is left free in the fit.


The number of expected events $\mu_k$ in Eq.~\ref{eq:likeli} is given by:

\begin{eqnarray}
\nonumber \mu_{k} &=& f_n \times \Bigg\{ 
\sum_{j} R_{CC\pi^0}(E_\nu^{j}) F_j\Big[S_{k}^{QE}(E_\nu^j)+\\
\nonumber &+& R_{res}S_{k}^{res}(E_\nu^j)+ R_{nres}S_{k}^{nres}(E_\nu^j) \Big]+ \\
&+&  B_{k}^{QE} + R_{res}B_{k}^{res} + R_{nres}B_{k}^{nres} + B_{k}^{NC}
\Bigg\}
 \label{eq:likeli-muk}
\end{eqnarray} 

\noindent  
$S_{k}$ and $B_{k}$ are the Monte Carlo events, respectively 
signal and background, contributing to the final 
CC$\pi^0$ sample in each bin $k$ of reconstructed neutrino energy.
The Monte Carlo signal events are 
further divided in 4 bins $E_\nu^j$ according to
their true neutrino energy: $0$-$1.5$ GeV, 
$1.5$-$2.0$ GeV, $2.0$-$2.5$ GeV and greater than $2.5$ GeV. 
The factors $F_j$ are defined as: 

\begin{eqnarray*}
F_j=\frac{ \sum_{k} \Big[S_{k}^{QE}(E_\nu^j)+
                S_{k}^{res}(E_\nu^j)+ S_{k}^{nres}(E_\nu^j) \Big]} 
           { \sum_{k} \Big[S_{k}^{QE}(E_\nu^j)+
           R_{res}S_{k}^{res}(E_\nu^j)+ R_{nres}S_{k}^{nres}(E_\nu^j) \Big]}
\end{eqnarray*}

\noindent 
in order to keep the normalization of the signal events independent from  
$R_{res}$ and $R_{nres}$. 

The fitting parameters are  $R_{CC\pi^0}(E_\nu^j)$ ($j=1,4$), $R_{res}$,
$R_{nres}$ and $f_n$.  The best fit of $R_{CC\pi^0}(E_\nu^j)$
gives the double ratio data over Monte Carlo 
between the number of inclusive CC$\pi^0$ events 
and the number of CCQE events, as a function of the true neutrino energy:

\begin{eqnarray}
 R_{CC\pi^0}(E_\nu^j) =\frac{N_{CC\pi^0}^{\mathrm true}(E_\nu^j)/
                                         N_{CCQE}^{\mathrm true}}
                      {N_{CC\pi^0}^{\mathrm MC}(E_\nu^j)/N_{CCQE}^{\mathrm MC}}
\label{eq:Rccpi0}
\end{eqnarray}

The scaling of the inclusive CC$\pi^0$ contribution in the fit is energy dependent
while the energy dependence of the CCQE is fixed to the Monte Carlo 
prediction since it has been accurately measured by previous experiments 
(\cite{Barish:1977qk,Bonetti:1977cs,Ciampolillo:1979wp,Belikov:1985kg}).
The corresponding uncertainty is considered a source of systematic error.
We also performed an energy independent fit of the CC$\pi^0$ to CCQE ratio,
following the same approach as Eq.~\ref{eq:likeli-muk} but with 
a single fit parameter $R_{CC\pi^0}$ rescaling
the CC$\pi^0$ contribution regardless of the true neutrino energy.

Table~\ref{tab:fit-all} shows the best fit values of
 $R_{CC\pi^0}$ for the energy-independent fit and the the 
four parameters $R_{CC\pi^0}(E_\nu^j)$ for the energy-dependent fit.

\begin{table}
\centering
\begin{tabular}{c|c}
\hline
\textbf{Fit Variable}&\textbf{Fit Result}\\
\hline \hline
\multicolumn{2}{c}{\textbf{Energy independent fit}}\\
\hline
$R_{CC\pi^0}$ & 1.436$\pm$0.109 \\
\hline
$R_{res}$ & 1.152$\pm$0.101 \\
\hline
$R_{nres}$ & 1.373$\pm$0.241 \\
\hline
$f_{n}$ & 0.968$\pm$0.025 \\
\hline \hline
\multicolumn{2}{c}{\textbf{Energy dependent fit}}\\
\hline
$R_{CC\pi^0}(E_\nu^1)$ & 1.005$\pm$0.027 \\
\hline
$R_{CC\pi^0}(E_\nu^2)$ & 1.180$\pm$0.127 \\
\hline
$R_{CC\pi^0}(E_\nu^3)$ & 1.307$\pm$0.198 \\
\hline
$R_{CC\pi^0}(E_\nu^4)$ & 1.418$\pm$0.129 \\
\hline
$R_{res}$ & 1.105$\pm$0.098 \\
\hline
$R_{nres}$ & 1.479$\pm$0.233 \\
\hline
$f_{n}$ & 0.980$\pm$0.021 \\
\hline \hline
\end{tabular}
\caption{Energy dependent and independent fit results 
for  $R_{CC\pi^0}(E_\nu^j)$. \label{tab:fit-all}}
\end{table}

The $\chi^2/{\mathrm d.o.f.}$ before the fit is $7135/44=162.1$.
The $\chi^2/{\mathrm d.o.f.}$ for the best fit 
is $40.2/37=1.095$ for the energy-dependent fit and 
$43.8/40=1.089$ for the energy independent fit.

The errors quoted for $R_{CC\pi^0}$ are purely statistical. 
The error induced on $R_{CC\pi^0}$ by the absolute normalization $f_n$ and by 
$R_{res}$ and $R_{nres}$ is evaluated in the fit in order to take in to account correlations but it is
considered a systematic error and reported in the first row of Table~\ref{tab:systqe}
together with the other sources of systematic in Sec.~\ref{sec:systematic}.

\begin{figure}[htbp!]
    \begin{center}
    \includegraphics[width=0.45\textwidth]{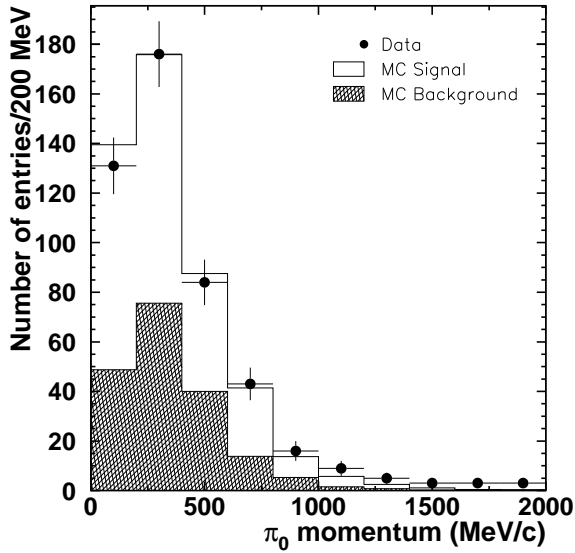}
    \caption{$\pi^0$ momentum distribution obtained rescaling the Monte Carlo with the energy independent fit result.}
    \label{fig:afterfitpmom}
    \end{center}
\end{figure}

\begin{figure}[htbp!]
    \begin{center}
    \includegraphics[width=0.45\textwidth]{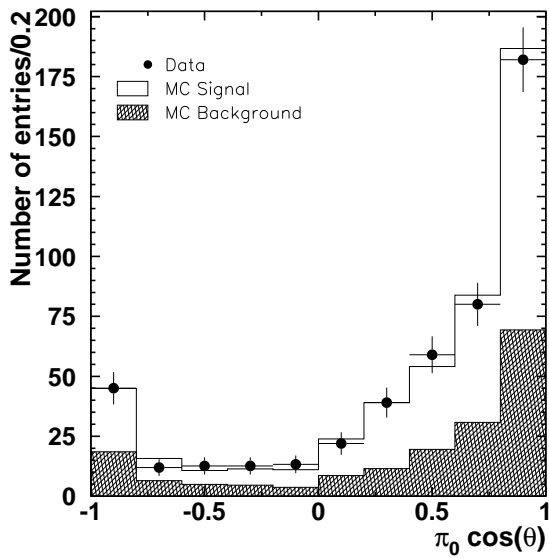}
    \caption{$\pi^0$ $cos(\theta)$ distribution obtained rescaling the Monte Carlo with the energy independent fit result.}
    \label{fig:afterfitcosth}
    \end{center}
\end{figure}

Figures~\ref{fig:afterfitpmom} and \ref{fig:afterfitcosth} show the
reconstructed $\pi^0$ momentum and angle with respect to the beam
direction in the laboratory frame, with the inclusive
CC$\pi^0$ production in the Monte Carlo rescaled to the best fit
value for both signal and backgrounds.

The fit results in Table~\ref{tab:fit-all} show an excess of CC$\pi^0$ production
with respect to our reference MC model. The energy dependent fit shows that the 
excess increases with the neutrino energy.
The data to MC ratio for the non resonant processes $R_{nres}$ is larger
than one while the ratio for the resonant contribution $R_{res}$ is consistent with one within
the statistical uncertainty only.
The resonant production with respect to the CCQE cross section
was measured by the K2K collaboration in the
CC1$\pi^+$ channel \cite{analisa} and found to be:
$0.734^{+0.140}_{-0.153}$, in very good
agreement with the MC prediction ($0.740\pm0.002(stat)$). 
According to our reference MC model, 50\% of the non resonant events have
one or more $\pi^0$ in the final state and 44\% of the selected
CC$\pi^0$ sample is produced in non resonant processes. Constraining the
resonance production to the experimental value and uncertainty given above,
we can use our CC$\pi^0$ sample to measure the non resonant contribution. 

We define CCnres (the CC non-resonant cross section) as the difference between the total
CC cross section and the sum of quasi elastic and resonance production.
Using the CC$\pi^0$ sample we perform an 
energy dependent and an energy independent fit of the CCnres to CCQE ratio,
following the same approach as Eq.~\ref{eq:likeli-muk}
but with $R_{CC\pi^0}$ and the normalization factors $F_j$ fixed to one. 
The fit parameter $R_{nres}$ rescaling the CC non resonant contribution in the energy 
independent fit (the parameters $R_{nres}(E_\nu^j$) in the energy dependent fit) 
and the overall normalization $f_{n}$ were left free in the fit.
The parameter $R_{res}$ was also free and the 
experimental constraint was incorporated in the fit by adding the term 
$\frac{(R_{res}-0.99)^2}{(0.21)^2} $ to the Log-likelihood function derived 
from Eq.~\ref{eq:likeli}.


Table~\ref{tab:fit-DIS} shows the best fit values for the 
energy independent fit of $R_{nres}$
and for the energy dependent fit of $R_{nres}(E_\nu^j)$.

\begin{table}
\centering
\begin{tabular}{c|c}
\hline
\textbf{Fit Variable}&\textbf{Fit Result}\\
\hline \hline
\multicolumn{2}{c}{\textbf{Energy independent fit}}\\
\hline
$R_{nres}$ & 1.461$\pm$0.118 \\
\hline
$R_{res}$ & 1.112$\pm$0.098 \\
\hline
$f_{n}$ & 1.030$\pm$0.012 \\
\hline \hline
\multicolumn{2}{c}{\textbf{Energy dependent fit}}\\
\hline
$R_{nres}(E_\nu^1)$ & 1.064$\pm$0.266 \\
\hline
$R_{nres}(E_\nu^2)$ & 0.872$\pm$0.252 \\
\hline
$R_{nres}(E_\nu^3)$ & 1.356$\pm$0.170 \\
\hline
$R_{nres}(E_\nu^4)$ & 1.567$\pm$0.164 \\
\hline
$R_{res}$ & 1.109$\pm$0.102 \\
\hline
$f_{n}$ & 1.026$\pm$0.012 \\
\hline \hline
\end{tabular}
\caption{Energy dependent and independent fit results 
for  $R_{nres}(E_\nu^j)$.\label{tab:fit-DIS}}
\end{table}

The $\chi^2/{\mathrm d.o.f.}$ for the best fit 
is $41.8/39=1.07$ for the energy dependent fit and 
$75.5/42=1.80$ for the energy independent fit. 
The value of the $\chi^2/{\mathrm d.o.f.}$ for the energy independent fit
shows that a three parameters fit of signal and background, not taking into 
account the energy dependence of the non-resonant contributions, gives a poor description of our data.


This result is obtained assuming that $\pi^0$ and $\pi^+$ production from
resonances are constrained by the same parameter within the Rein-Sehgal
model.  As a consistency check we repeated the fit without constraint to
the CC$\pi^+$ measurement and the results are consistent within 3\% with the
results in Table~\ref{tab:fit-DIS}. This difference is accounted by the systematic
due to the uncertainty in the non-QE composition evaluated in Sec.~\ref{sec:systematic}.

The errors quoted for $R_{nres}$ are purely statistical. 
The error induced on $R_{nres}$ by the normalization $f_n$ and by $R_{res}$ 
is evaluated in the fit in order to take in to account correlations 
but it is considered a systematic error and reported in the first row of Table~\ref{tab:systqe}.
The full systematic error is evaluated in Sec.~\ref{sec:systematic}. 

\section{\label{sec:systematic}Systematic error study}
In this section we discuss the sources of systematic error. 
The contributors to the systematic error
on the energy independent results CC$\pi^0$ and CCnres are summarized 
in Table~\ref{tab:systqe}.
The systematic errors for the energy dependent results in
Tables~\ref{tab:measurement-pi0all} and \ref{tab:CCDIS-results} are calculated
following the same approach.

\begin{table}[htbp!]
\centering 
\begin{tabular}
    {c|c|c} \hline
    \textbf{Source}&\textbf{$\dfrac{\sigma_{CC\pi^{0}}}{\sigma_{CCQE}}[\%]$} & \textbf{$\dfrac{\sigma_{CCnres}}{\sigma_{CCQE}}[\%]$} \\ \hline \hline
    Normalization and fit &                     -1.8 +1.8 & -3.5 +3.5 \\ \hline
    non-QE CC cross sections &                        -3.4 +3.3 & -3.1 +3.7 \\ \hline
    Bodek\&Yang corr. &                         -4.3 +3.5 & -8.3 +7.8 \\ \hline
    CCQE $M_A$           &                              -1.3 +2.4 & -0.8 +1.4 \\ \hline
    NC/CC ratio &                                       -0.5 +0.5 & -0.8 +0.8 \\ \hline
    $\nu$ flux &                                     -0.1 +0.1 & -0.4 +0.4 \\ \hline
     $E_{\nu}$ reco. parameters &                                     +0.2 -0.2 & +0.3 -0.3 \\ \hline
    \textbf{Int. model/flux} & \textbf{ -5.9 +5.7} & \textbf{-9.6 +9.5}\\ \hline
    $\pi$ absorbtion &                    -2.0 +2.1 & -1.8 +2.0 \\ \hline
    $\pi$ inelastic &                     -3.0 +1.8 & -2.2 +1.5 \\ \hline
    Proton rescattering &                               -1.9 +0.3 & -2.8 +2.6 \\ \hline
    Pion inter. length &                           -1.5 +1.5 & -2.9 +2.3 \\ \hline
    \textbf{Nucl. model} & \textbf{                -4.1 +2.8} &\textbf{-4.9 +4.3} \\ \hline
    PMT resolution &                                    -0.5 +0.1 & -0.6 +0.7 \\ \hline
    Scintillator quench. &                            -0.1 +0.5 & -0.4 +0.5 \\ \hline
    Cross-talk &                                        +1.2 +2.6 & +1.4 +2.3 \\ \hline
    PMT threshold &                                     -1.7 +2.0 & -1.6 +2.2 \\ \hline
    \textbf{Detector effects} & \textbf{                -1.8 +3.3} & \textbf{-2.2 +3.5}\\ \hline
    Fiducial Volume  &                                  -2.6 +2.5 & -3.3 +3.2\\ \hline
    Vertex disc.                    &          -1.9 +2.4 & -3.2 +2.6\\ \hline
    Vertex pointing &                                   -1.0 +1.8 & -1.4 +1.9\\ \hline
    EC cluster energy &                             -0.4 +1.2 & -0.4 +1.1\\ \hline
    \textbf{Selection cuts} & \textbf{                  -3.4 +4.1} & \textbf{-4.8 +4.7}\\ \hline\hline
    \textbf{Total} & \textbf{                           -8.1 +8.2} &\textbf{-12.0 +12.0} \\
    \hline
\end{tabular}
\caption{{Systematic errors for the CC$\pi^0$ and the CC non resonant cross
sections, relative to the CCQE cross section.}}
\label{tab:systqe}
\end{table}

\subsection{Interaction model and neutrino flux}

The error coming from the absolute normalization for $R_{CC\pi^0}$
is estimated repeating the corresponding fit while fixing 
all the other parameters including the absolute normalization $f_n$ 
at their best fit values. The resulting error for $R_{CC\pi^0}$ 
is the one reported in the previous section 
as pure statistical error while its quadratic difference with the full fit error
is reported as normalization errors in Table~\ref{tab:systqe}. The same procedure
is applied to evaluate the errors on $R_{nres}$.

The systematic error coming from the uncertainty in the
composition of the non-QE CC cross section has been taken into account 
assigning a weight factor $W_i$ to each non-QE CC channel and repeating the fit.
A constraining function
$F_{syst}$ was added to the Log-likelihood derived from Eq.~\ref{eq:likeli}:

\begin{eqnarray}
F_{syst}=\frac{\sum_i(W_i-1)^2}{\sigma_{W^2}}
\label{eq:fsyst}
\end{eqnarray}

\noindent where $\sigma_{W}=30\%$.
The double ratio data to Monte Carlo of the NC to CC 
processes is also left free in the fit with a $20\%$
constraint, adding the corresponding term to $F_{syst}$ described
in Eq.~\ref{eq:fsyst}. 
The total systematic error listed in the second row of Table~\ref{tab:systqe} takes
into account the correlation between the different sources above.

The QE axial mass is varied by $\pm10\%$ (according to the
uncertainties in the measurement reported in \cite{Gran:2006jn}) and the
Bodek and Yang corrections to DIS events by $\pm$30$\%$\cite{Bodek:2002vp}. The
resulting systematic errors are added in quadrature in the total
interaction model uncertainty. 

The uncertainty in the shape of the
neutrino energy spectrum shown in Fig.~\ref{fig:spectrum} is
considered by changing the flux in each bin, taking into account
their errors and the correlations between them (see Ref.~\cite{Ahn:2006zz}
for details).

The uncertainty on the values assumed for the neutrino energy
reconstruction parameters in Eq.~\ref{eq:neutrinoenergy} is evaluated by changing 
the value of $W$=1.483 GeV by $\pm$15\%, corresponding to assume that all non-qe selected events 
are from $\Delta$(1232) rather than from an average $W$=1.483GeV
The nuclear potential $V$ is varied from 0 to 27 MeV.
%

%
\subsection{Nuclear model}
Nuclear effects alter the composition and kinematics of the
particles produced in neutrino interactions in nuclei. 

Pion absorption and inelastic scattering processes, in particular
pion charge exchange, modify the $\pi^0$ yield. To account for
the uncertainty in the Monte Carlo modeling of these effects, the
pion absorption and pion inelastic scattering cross sections are
varied by $\pm30\%$ \cite{Ingram:1983}. The proton rescattering is changed by
$\pm10\%$ according to the
uncertainties derived from cross section measurements
\cite{jeon:2003,Ingram:1983}. The systematic errors in
Table~\ref{tab:systqe} are calculated by repeating the analysis for each
variation of the corresponding source. The uncertainty in the pion
interaction length is considered by changing its value by $\pm20\%$.
The overall uncertainty on the MC model is calculated considering
the uncertainty in the pion interaction length fully correlated to
the pion inelastic cross section above.
\subsection{Detector effects}
The SciBar hit threshold, set at 2 photo-electrons, is changed by
$\pm30\%$ and the corresponding variation of the result is quoted as
a systematic error. 

The model for the cross-talk in SciBar takes into
account the second neighboring pixel and has a single free parameter
$\textit{n}$ corresponding to the fraction of charge given by
cross-talk in the adjacent pixel. The best fit obtained comparing
data and Monte Carlo is $\textit{n}=(3.25\pm0.01)\times10^{-2}$. The
same model is used for the crosstalk simulation in Monte Carlo and
for the correction of the crosstalk effect both in Monte Carlo and
data. To evaluate the systematic due to the crosstalk, we changed
the crosstalk parameter $\textit{n}$ in the simulation in the range
from 3.0\% to 3.5\%, corresponding to the uncertainty in the
crosstalk modeling\cite{analisa}

Smaller systematic detector effects are induced
by the uncertainties in the single photo-electron PMT resolution and
the scintillator quenching (Birk's saturation) in SciBar. The SciBar
PMTs resolution in the Monte Carlo is set at 40\%\cite{Hasegawa-thesis}. This value was
chosen by tuning the dE/dx per plane for muons in Monte Carlo to
match the response to cosmic ray data. The uncertainty is evaluated
to be 10\% and the corresponding systematic errors are listed in
Table~\ref{tab:systqe}. The scintillator quenching in SciBar was
measured in a beam test and is well reproduced by Birk's equation\cite{Hasegawa-thesis}.
The systematic error is evaluated by varying the Birk's parameter
within its uncertainty. Other detector effects were found to give
negligible contributions to the systematic error.
\subsection{Selection cuts}
In order to evaluate the systematic uncertainties due to the selection, in 
Table~\ref{tab:systqe} we quote the dependence of the result to
variations of the cuts. 

We change the fiducial volume by changing
(simultaneously in data and Monte Carlo) one at the time the
fiducial volume cuts in the three coordinates according to the
resolution for reconstructing the neutrino vertex: 0.8 cm for both X
and Y, and 1.6 cm for the Z. Then we add in quadrature the three
corresponding variations of the result. 

The systematic uncertainty on the cut requiring the photon track to
be disconnected from the
vertex has been assessed by looking at the resolution on
the neutrino vertex reconstruction and adding in quadrature the
resolution on the photon conversion point. We assumed the 
resolution on the photon conversion point to be equal
to the resolution on the muon vertex. 

The cut on the  photon track pointing to the vertex 
is applied to the distance by which the photon candidate track
misses the vertex when extrapolated to the vertex plane. We take
0.14 rad for the 2D angle resolution of photon direction (0.12 rad
and 0.14 rad, respectively, for the most energetic photon and the least
one). The cut is applied on photons disconnected by more than 20 cm
from the vertex. They have on average a 50 cm distance from the vertex.
We set the variation as $50\cdot0.14=7.0$ cm. In the cut region (25
cm), agreement between MC and data is quite satisfactory. 

\section{\label{sec:results}Results}
\begin{figure}[htbp!]
  \begin{center}
    \includegraphics[width=0.45\textwidth]{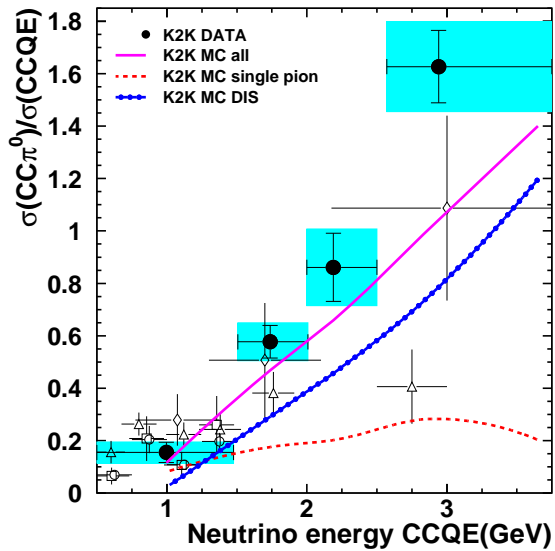}
    \caption{CC$\pi^0$ to CCQE cross section ratio  as a function of neutrino energy.
    The result of this analysis ($\bullet$) is compared with our standard MC expectation 
    and past experimental results. 
    The previous experimental data are:
    ($\bigcirc$)ANL82\protect\cite{Radecky:1982fn},($\Box$)ANL\protect\cite{Barish:1977qk},
    ($\triangle$)BNL86\protect\cite{Kitagaki:1986ct},($\Diamond$)ANL83+BNL86\protect\cite{Day}\cite{Kitagaki:1986ct}(see text for details).
    }
    \label{fig:CCcrosscompdm}
    \end{center}
\end{figure}

Under the assumption that the detection efficiency are the same in data and Monte Carlo, 
the ratio between the inclusive CC$\pi^0$ 
cross section and the CCQE cross section
can be calculated from Eq.~\ref{eq:Rccpi0} multiplying the best fit values of 
$R_{CC\pi^0}$ given in Table~\ref{tab:fit-all} by the MC
prediction for the cross section ratio in each neutrino bin $E_\nu^j$:
\begin{eqnarray}
  \frac{\sigma_{CC\pi^0}}{\sigma_{CCQE}}(E_\nu^j)= 
      R_{CC\pi^0}(E_\nu^j) \times  
       \left[\frac{\sigma_{CC\pi^0}}{\sigma_{CCQE}}(E_\nu^j)\right]_{\mathrm MC}
    \label{eq:sigmaCCpi0}
\end{eqnarray}
%
%
Table~\ref{tab:measurement-pi0all}
shows the CC$\pi^0$ cross section ratio to CCQE,  
integrated over all energies and as a function of the four neutrino energy bins.
Figure~\ref{fig:CCcrosscompdm} shows the result as a function of neutrino energy.
The vertical bars are the statistical errors, the height of the filled areas 
corresponds to the statistical and systematic errors added in quadrature and 
the data points in each bin are set at the weighted averages of the true neutrino
energy for the selected CC$\pi^0$ events.
Figure~\ref{fig:CCcrosscompdm} also shows the CC$\pi^0$ over CCQE ratio and the 
two largest contributions, single pions from resonances and pions produced in DIS,
as they are predicted by our reference MC. According to our reference MC the average 
true neutrino energy for the selected CCQE events is 1.1 GeV. The average true 
neutrino energy for our selected $\pi^0$ sample is 1.3 GeV and 2.5 GeV for the fraction 
of $\pi^0$ produced in DIS events.

\begin{table}
\centering
\begin{tabular}{c|c}
\hline \hline
\textbf{Energy Range}&\textbf{Cross Section Ratio}\\
\textbf{GeV}&$\dfrac{\sigma_{CC\pi^{\circ}}}{\sigma_{CCQE}}$\\
\hline \hline
$>$ 0.0 & 0.426 $\pm$ 0.032(stat.)$\pm$ 0.035(\text{syst.}) \\
\hline \hline
0.0 - 1.5 &0.155 $\pm$ 0.039(stat.)$\pm$0.010(\text{syst.})\\
\hline
1.5 - 2.0 &0.577 $\pm$ 0.062(stat.)$\pm$0.037(\text{syst.})\\
\hline
2.0 - 2.5 &0.861 $\pm$ 0.130(stat.)$\pm$0.067(\text{syst.})\\
\hline
$\geq$ 2.5 &1.627 $\pm$ 0.138(stat.)$\pm$0.103(\text{syst.})\\
\hline \hline
\end{tabular}
\caption{Inclusive Cross Section Ratio
$\dfrac{\sigma_{CC_{\pi^\circ}}}{\sigma_{CCQE}}$ as a function of the
neutrino energy.} \label{tab:measurement-pi0all}
\end{table}

The CC non resonant to CCQE ratio is obtained from the 
best fit values of $R_{nres}$ in Table ~\ref{tab:fit-DIS} similarly to 
Eq.~\ref{eq:sigmaCCpi0}:
\begin{eqnarray}
\frac{\sigma_{CCnres}}{\sigma_{CCQE}}(E_\nu^j)= 
      R_{nres}(E_\nu^j) \times  
       \left[\frac{\sigma_{CCnres}}{\sigma_{CCQE}}(E_\nu^j)\right]_{\mathrm MC}
\end{eqnarray} 

The results for $\frac{\sigma_{CCnres}}{\sigma_{CCQE}}$ are reported in Table~
\ref{tab:CCDIS-results} integrated over all energies and as a function of the 
four neutrino energy bins.

\begin{table}
\centering
\begin{tabular}{c|c}
\hline \hline
\textbf{Energy Range}&\textbf{Cross Section Ratio}\\
\textbf{GeV}&$\dfrac{\sigma_{CCnres}}{\sigma_{CCQE}}$\\
\hline \hline
$>$ 0.0 & 0.419 $\pm$ 0.034(stat.)$\pm$ 0.050(\text{syst.}) \\
\hline \hline
0.0 - 1.5 & 0.010 $\pm$ 0.002(stat.)$\pm$0.002(\text{syst.})\\
\hline
1.5 - 2.0 & 0.432 $\pm$ 0.125(stat.)$\pm$0.056(\text{syst.})\\
\hline
2.0 - 2.5 & 1.304 $\pm$ 0.164(stat.)$\pm$0.117(\text{syst.})\\
\hline
$\geq$ 2.5 & 2.954 $\pm$ 0.309(stat.)$\pm$0.354(\text{syst.})\\
\hline \hline
\end{tabular}
\caption{Cross Section Ratio $\dfrac{\sigma_{CCnres}}{\sigma_{CCQE}}$ as a function of the
neutrino energy.} \label{tab:CCDIS-results}
\end{table}

\subsection{\label{sec:Comparisonwithotherexperiment} Comparison with
other experiments}
Past experimental results exist for the exclusive $\nu_{\mu}n
\rightarrow \mu^- p \pi^0$ cross sections on deuterium (Barish
\cite{Barish:1978pj}, Radecky \cite{Radecky:1982fn} and Kitagaki
\cite{Kitagaki:1986ct}). There is also a published result for the
exclusive cross section $\nu+p\rightarrow\mu^-p\pi^+\pi^0$
(Day \cite{Day}). In order to compare with our result on C$_8$H$_8$,
cross sections on deuterium have been rescaled to the
different number of protons and neutrons. The ratio between
CC$\pi^0$ and CCQE cross sections is computed by
dividing the experimental results quoted above by the CCQE cross
section measured by Barish \cite{Barish:1977qk}. Below 1.5 GeV neutrino energy our result can be directly compared
with the published single pion cross sections, since this is the main
contribution to the inclusive cross section. 
The three points
shown as diamond-shaped symbols at 1.07 GeV, 1.70 GeV and 3.0 GeV are obtained adding the two pion
$\mu^- p \pi^+ \pi^0$ from \cite{Day} to the single pions $\mu^- p
\pi^0$ taken from \cite{Kitagaki:1986ct}. 

The MiniBooNE collaboration has recently published charged current $\pi^0$ production~\cite{AguilarArevalo:2010xt}
at a slightly lower neutrino energy (between 0.5 GeV and 2 GeV) and their result is consistent with
the one presented in this paper.
\section{\label{sec:Conclusion} Conclusion}
Out of a sample of 11,606 charged-current neutrino interactions in
the K2K SciBar detector, we selected 479 inclusive $\pi^0$ events
with an efficiency of 7.8\% and a purity of 66.5\%. The sample
corresponds to $2.02\times10^{19}$ protons on target recorded with
the SciBar+EC+MRD detectors at 1.3 GeV average beam neutrino energy. 
We measured the cross section for charged-current $\pi^0$ production
relative to the charged-current quasi-elastic cross section to avoid the large 
uncertainties in the absolute neutrino flux determination. 
The result integrated over the neutrino energy spectrum is
\begin{eqnarray*}
\frac{\sigma_{CC{\pi^0}}}{\sigma_{CCQE}}=0.426\pm0.032(stat)\pm
0.035(syst.)
\end{eqnarray*}
higher than the prediction 
of our reference Monte Carlo.
The energy dependent CC$\pi^0$ to CCQE cross section ratio 
is presented in Table~\ref{tab:measurement-pi0all} and shown in Fig.~\ref{fig:CCcrosscompdm}.
The results of the best fit for the composition of our CC$\pi^0$ sample
show that the data excess comes from non resonant processes, mainly
$\pi^0$ production in DIS, rather than from $\pi^0$ in resonance production.
Using the measured CC single charged pion cross section \cite{analisa} 
as a constraint for the resonant production, 
we measured the ratio between the CC non resonant and the CCQE cross section,
integrated over the neutrino energy spectrum:
\begin{eqnarray*}
\frac{\sigma_{CCnres}}{\sigma_{CCQE}}=0.419\pm0.034(stat.)\pm0.050(syst.)
\label{eq:mpiqecross}
\end{eqnarray*}
For CC non resonant processes we define any charged current process
except quasi elastic interaction and resonances production.
The energy dependent cross section ratio is presented in 
Table~\ref{tab:CCDIS-results}.
The results presented here are the firsts for neutrinos of few-GeV energy 
on a C$_8$H$_8$ target material and improve the precision of previous results on
different targets and therefore are a significant 
contribution to the knowledge of neutrino interaction processes
relevant for several present and future oscillation experiments.

\section{Acknowledgment}

We thank the KEK and ICRR directorates for their
strong support and encouragement. K2K was made possible
by the inventiveness and the diligent efforts of the
KEK-PS machine group and beam channel group. We
gratefully acknowledge the cooperation of the Kamioka
Mining and Smelting Company. This work has been supported
by the Ministry of Education, Culture, Sports,
Science and Technology of the Government of Japan,
the Japan Society for Promotion of Science, the U.S.
Department of Energy, the Korea Research Foundation,
the Korea Science and Engineering Foundation, NSERC
Canada and Canada Foundation for Innovation, the Istituto
Nazionale di Fisica Nucleare (Italy), the Ministerio
de Educaci\`{o}n y Ciencia and Generalitat Valenciana
(Spain), the Commissariat \`{a} l'Energie Atomique
(France), and Polish KBN grants: 1P03B08227 and
1P03B03826.

\bibliographystyle{unsrt}
\bibliography{references}

\end{document}